\documentclass{article}
\usepackage{spconf,amsmath,graphicx,url}
\usepackage{comment}

\newcommand{\aug}{Kaldi}

\title{Acoustic modeling for Overlapping Speech Recognition:\\
JHU Chime-5 Challenge System}
\name{Vimal Manohar$^{1,2}$, Szu-Jui Chen$^1$, 
  Zhiqi Wang$^1$, Yusuke Fujita$^{3,1}$, Shinji Watanabe$^{1}$, Sanjeev Khudanpur$^{1,2}$
\thanks{This work was partially supported by NSF Grant No CRI-1513128 and
IARPA MATERIAL award number FA8650-17-C-9115. Vimal Manohar was supported by Alexa Graduate Fellowship.}}
\address{
  $^1$Center for Language and Speech Processing\\
  $^2$Human Language Technology Center Of Excellence\\
  Johns Hopkins University, Baltimore, MD 21218 \\
  $^3$Hitachi, Ltd. Research \& Development Group, Kokubunji-shi, Tokyo, Japan \\
  \texttt{vimal.manohar91@gmail.com,\{schen146,zwang132,shinjiw,khudanpur\}@jhu.edu}}
\begin{document}
%
\maketitle
\begin{abstract}
This paper summarizes our acoustic modeling efforts in the Johns Hopkins University speech recognition system for the CHiME-5 challenge to recognize highly-overlapped dinner party speech recorded by multiple microphone arrays. 
We explore data augmentation approaches, neural network architectures, front-end speech dereverberation, beamforming and robust i-vector extraction with comparisons of our in-house implementations and publicly available tools.
We finally achieved a word error rate of 69.4\% on the development set, which is a 11.7\% absolute improvement over the previous baseline of 81.1\%, and release this improved baseline with refined techniques/tools as an advanced CHiME-5 recipe. 
\end{abstract}
\section{Introduction}
\label{sec:intro}

Automatic speech recognition (ASR) has improved significantly 
in the area of conversational speech recognition. A lot of this 
can be attributed to larger datasets for training, 
new sequence training objectives like
Connectionist Temporal Classification (CTC) \cite{graves2006ctc} 
and Lattice-free MMI (LF-MMI) \cite{chain},
and improved neural network architectures for acoustic modeling. 
Recently, super-human performance has been achieved on conversational 
telephone speech 
using publicly available datasets \cite{kurata2017highwaylstm,han2017capio,xiong2018microsoft}. 
However, speech recognition in more difficult settings like reverberant, noisy, and overlap speech conditions, still lags behind, and is widely believed to be the next frontier in speech recognition. 
Interest in robust and distant speech recognition has increased greatly with new voice control applications in home environments, cars, etc.

The CHiME-5 challenge \cite{barker2018chime5} focuses on the problem of distant microphone 
conversational speech recognition in the setting of everyday home environments. 
The speech data consists of real 4-people dinner party conversations recorded
using linear array microphones, and annotated with speech start/end times,
the speaker labels and speech transcription.

The challenge organizers developed the baseline system using state-of-the-art techniques including LF-MMI and time-delay neural network (TDNN) architecture \cite{barker2018chime5}.
However, the word error rate (WER) of the baseline system was 81.1\% and did not reach a satisfied level.
This huge number for WER implies two challenging problems in the CHiME-5.
One challenging problem is that distant microphones are distributed in different home locations and speakers are not always facing to any microphones.
Another big problem is spontaneous and overlapping nature of speech -- 
speakers in the parties are all friends and instructed to behave naturally.

First, to tackle the distributed distant microphone setting, we explore data augmentation for improving robustness against a variety of room acoustics and directivity patterns of speech, which shows great success in noise robust ASR \cite{kinoshita2013reverb,harper2015automatic,vincent2017analysis}.
We also explore neural network architectures for 
capturing the variation introduced by data augmentation.
In addition, the effectiveness of dereverberation techniques in different network architectures is evaluated.
Second, to tackle spontaneous and overlapping nature of speech, a 2-stage decoding method is examined. This method extracts reliable single-speaker frames for i-vector extraction \cite{vijayaditya2015ami}.

While we developed an advanced ASR system with other colleagues as the Hitachi/JHU CHiME-5 Challenge submission \cite{kanda2018hitachi} during the challenge period, we used different implementations in this work.
One of the most important contributions of this work is to provide our outcomes in a publicly available new Kaldi recipe\footnote{https://github.com/kaldi-asr/kaldi/tree/master/egs/chime5/s5b} so that the researchers in this field can reproduce this new advanced baseline.
We explore the implementation of above mentioned techniques using publicly available tools including Nara WPE \cite{drude2018nara-wpe} and RIR generator \cite{habets2006rir}.
We include these public tools in our Kaldi recipe to build a single-system baseline not using any system combination. 
%
%



\section{CHiME-5 Conventional ASR System}
\label{sec:chime5-baseline-system}

This section describes the CHiME-5 data and the baseline ASR system released by the challenge organizers \cite{barker2018chime5}.

\subsection{CHiME-5 Data}
The dataset consists of 20 dinner parties, each recorded with six 
Kinect\footnote{Kinect is a registered trademark of Microsoft Corporation. } devices placed in different locations. Each Kinect device 
has a linear array of 4 sample-synchronized microphones. 
16 of these parties form the training set, 2 of them are in the development set, and the rest are in the evaluation set.
Since the same speech is recorded from multiple 
locations and channels, we randomly select some of the 
utterances for training. We refer to these subsets of audio based on the amount
of utterances selected e.g. ``100k'', ``400k''.
In addition, each speaker is also recorded using a set of worn binaural 
microphones to provide parallel (relatively) ``clean" speech. 
We arbitrarily selected to use only the left channel from this.
We refer to this set of audio as ``worn'' data.

\subsection{HMM-GMM System}
A HMM-GMM system is used as a seed system to get alignments for 
neural network training. We use the same approach as in \cite{barker2018chime5} 
for training this system; however we use larger amounts of 
utterances from the array microphone data. While the baseline system
used ``worn + 100k" set for training, 
we get improved results using ``worn + 400k" set.
In this work, we use the same array synchronization and speech 
enhancement methods (only at test time), as well as the same lexicon and language models used in the original baseline system.

We also do data cleanup as in \cite{barker2018chime5,vijayaditya2015ami} to remove
irregularities in the utterances such as transcription errors,
excessive silence and noise, and train a new HMM-GMM system 
using the ``cleaned" version of ``worn + 100k" 
or ``worn + 400k" set.

\subsection{Neural network acoustic model}
The baseline system in \cite{barker2018chime5} 
used a simple time-delay neural network (TDNN). We 
make several improvements to this acoustic model as described in
the following section.

\section{Acoustic Modeling improvements}
\label{sec:am-improvements}

\subsection{Data Augmentation}
\label{sec:data-augmentation}
Simulating reverberated speech from worn microphone recordings is beneficial to improving robustness.
This simulation is performed by utilizing room impulse responses (RIR) and point-source noises.
We examine two approaches to generate RIRs to obtain the simulated recording.

\subsubsection{CHiME-5 data augmentation}
\label{ssec:chime5-data-augmentation}

In the first method, we use the image method proposed by Allen and Berkeley in \cite{imagemethod} to generate the RIR for each recording.
Here, we use approximated room size and the position of the microphones in reference to the provided floorplans.
All rooms are considered as a cube.
Then we randomly put a location for each speaker inside the room. The reverberation time is randomly sampled from a Gaussian distribution between 0 to 0.9 second. After getting all these configurations, we 
generate an RIR for each microphone.

Point-source noises are collected from the training data because we are not allowed to use external noise corpora in the CHiME-5 challenge.
We simply extract all the non-speech parts from the original recordings of CHiME-5 training data based on the annotated start/end times, 
and split them into 20-second chunks as noise chunks. 

Then, we randomly pick up 0 to 3 point-source noises assigned with a corresponding random position inside the room.
Each noise comprises of randomly selected 20-second noise chunks to make the noise have the same length as the reverberated speech.
The noise is convolved with the 
corresponding RIR generated to get reverberated noise. 
Finally, we mix it with the reverberated speech with a signal-to-noise ratio (SNR) 
randomly selected from 0 to 30dB, and normalize the audio with the same intensity as the original recording.

\subsubsection{{\aug} data augmentation}
\label{ssec:ko-data-augmentation}

Ko et. al \cite{ko2017augmentation} found that augmentation 
using a large number of synthetic RIRs can show improvements over using
even real RIRs. The synthetic RIRs were
generated using the RIR generator by Habets et. al \cite{habets2006rir}.
Ko et. al used additive noises from the 
MUSAN \cite{musan} corpus. 
This approach is the standard in many Kaldi \cite{kaldi} recipes, and we refer to this as the 
``{\em{\aug} data augmentation}".
In this paper, we compare this method 
with our CHiME-5 data augmentation described in Section \ref{ssec:chime5-data-augmentation}. 
Since the CHiME-5 challenge does 
not allow using external noise corpus, we a variation of this method
using the noises extracted from CHiME-5 data as described in Section \ref{ssec:chime5-data-augmentation}. 
Because of the simplicity of the latter method,
we include it in our new improved Kaldi recipe.

\subsection{Robust i-vector extraction}
\label{ssec:ivectors}

We use a 2-stage decoding approach used in \cite{peddinti2015aspire} to 
get i-vectors from reliable speech segments. We first perform a first-pass 
decode of the utterances using i-vectors extracted from the entire 
utterance. 
Only the regions 
corresponding to the real words that are recognized with a lattice posterior 
probability of 1.0 and a duration less than 1s are considered 
reliable for the purpose of i-vector extraction\footnote{We used the parameters that worked the best in \cite{peddinti2015aspire} and did not tune them for this work.}. This will exclude 
all the regions of silence and filler words, and ideally regions of 
overlapping speech, which is expected to have higher confusability (and hence
lower lattice posterior probability). A second-pass decoding is then done 
using i-vectors extracted using statistics obtained from only these reliable 
regions. This approach
gives a consistent 1.5-2\% absolute improvement of WER as shown in 
Section \ref{ssec:2stage-results}.

\subsection{Neural Network Architectures}
\label{ssec:nnet_architectures}

This section describes the various neural network architectures we 
explored for acoustic modeling -- TDNN, TDNN + LSTM, TDNN-F and CNN + TDNN + LSTM. 
We use batch normalization \cite{ioffe2015batchnorm} in all the TDNN and
CNN layers, and per-frame dropout in the LSTM layers
\cite{srivastava2014dropout,cheng2017dropout} with scheduling of dropout factor 
as suggested in \cite{cheng2017dropout}.

\subsubsection{TDNN}
TDNN is the architecture used in the baseline system. This consists of 
8 time-delay neural network (TDNN) layers of 512 hidden units 
and an additional TDNN layer before the output of the same size but factorized with a bottleneck of size 320.

\subsubsection{TDNN + LSTM}
TDNN + LSTM architecture consists of 4 long-short term memory with projection (LSTMP)
\cite{sak2014lstmp} 
layers of size 1024 and output and recurrent projection dimensions of 256.
The LSTMP layers are interleaved with TDNN layers. There are two TDNN layers
between a pair of LSTMP layers.

\newcommand{\W}{\mathbf{W}}
\newcommand{\A}{\mathbf{A}}
\newcommand{\B}{\mathbf{B}}

\subsubsection{TDNN-F}
TDNN-F architecture is a factored form of TDNN introduced in
\cite{povey2018tdnnf}. 
The basic factorization idea here is to
factorize the weights matrix $\W$ (of size $m \times n$) of a TDNN layer
as a product of two matrices: $\W = \A\B$, where 
$\A$ is of size $m \times p$ and $\B$ is of size $p \times n$ such that $p << m,n$ and matrix $\B$ is constrained to be semi-orthogonal. 

In our work, we use an architecture with 15 TDNN-F layers with a hidden dimension $n = 1536$ and a bottleneck dimension $p=160$. Each layer also has a resnet style 
bypass-connection from the previous layer's output i.e. the previous layer's
output (scaled by a factor of 0.66) is added to the current layer's output. 
The TDNN-F layers also use a ``continous dropout'' with scheduling as 
suggested in \cite{povey2018tdnnf}.

\subsubsection{CNN + TDNN + LSTM}
This architecture has in the beginning of the network two 2-D convolution layers
with 3x3 kernels with 256 and 128 filters respectively. 
There is a sub-sampling along the frequency dimension by a factor of 2 after the
first convolution. 
The convolution layers are followed by 3 LSTMP layers interleaved with two TDNN
layers between each pair of LSTMP layers. 

\section{Results}
\label{sec:results}

In this section, we report experimental results using the 
acoustic modeling improvements described in the previous section.
We train our neural networks with LF-MMI objective using the Kaldi 
toolkit \cite{kaldi}. For training details, the reader is 
directed to \cite{chain}. 
The neural networks use 40-dim MFCC features as input. I-vectors are used for speaker adaptation \cite{dehak2011ivectors,karafiat-2011-ivector-adaptation} with the i-vectors being extracted on top of PCA-reduced spliced-MFCC features. 
As in \cite{chain}, we always apply 3x speed perturbation of data and 
volume perturbation by a random factor in $[0.8,2.0]$ to improve
robustness of the models. 

\subsection{Effect of training data}
\label{ssec:training-data-results}

We compare using different amounts of training data and data augmentation methods
in Table \ref{tab:training-data}. Rows 1 and 2 in Table \ref{tab:training-data} 
show results when the ASR systems are trained using
a mix of the ``worn" data and array microphone data. In row 1, we use
100k utterances corresponding to $\sim$ 70 hours after cleanup
and in row 2, we use 400k utterances corresponding to $\sim$ 160 hours after cleanup. Note 
that this is not an increase in the real amount of data, but just involves using 
the same speech samples recorded using different microphone channels and locations. Since,
using 400k utterances gives a 3\% absolute improvement, for the rest of the 
experiments we use these many utterances. Increasing the number of utterances
to 800k did not result in any improvements in our experiments. 

Row 3 in Table \ref{tab:training-data} shows the results using the same system
as row 2, but applying dereverberation at test time. 
This gives a small improvement. In the rest of the experiments in this section, we apply
dereverberation at test time. Row 4 shows results using {\em{\aug} data augmentation} described in Section \ref{ssec:ko-data-augmentation} with additive noises from MUSAN
\cite{musan} corpus. We create 2 copies of corrupted ``worn" data
and mix it with microphone data. The Row 5 uses CHiME-5 data augmentation described in Section \ref{ssec:chime5-data-augmentation} with RIRs synthesized based on CHiME-5 room
conditions and noises extracted from the CHiME-5 training data. 
Row 6 uses {\em{\aug} data augmentation}, but using noises 
from CHiME-5 data. The results show that the latter method gives competing results. Because of its
simplicity and use of open-source tools, we include this augmentation method
in our new improved Kaldi recipe.

\begin{table}[htb]
  \caption{\label{tab:training-data}WER(\%) results for different training data}
  \centering
  \begin{tabular}{l|l|c|c|c}
    WPE & Training data & \multicolumn{2}{c|}{Data augmentation} &
    WER(\%) \\
    && RIRs & Noises \\
    \hline
    \hline
    No & worn + 100k & - & - & 81.3 \\
    No & worn + 400k & - & - & 78.8 \\
    Nara & worn + 400k & - & - & 78.2 \\
    \hline
    Nara & worn + 400k & Ko \cite{ko2017augmentation} & MUSAN & 75.4 \\
    Nara & worn + 400k & CHiME-5 & CHiME-5 & 75.3 \\
    Nara & worn + 400k & Ko \cite{ko2017augmentation} & CHiME-5 & \bf 73.9 \\
  \end{tabular}
\end{table}

\subsection{Effect of speech dereverberation}

In this section, we investigate the effect of applying weighted prediction
error (WPE) \cite{nakatani2010wpe} based dereverberation before beamforming at 
test time. We tried two different implementations of the WPE algorithm -- NTT WPE \cite{nakatani2010wpe}\footnote{\url{http://www.kecl.ntt.co.jp/icl/signal/wpe/}}, which has limited use based on NTT's strict licence, and
Nara WPE \cite{drude2018nara-wpe}\footnote{\url{https://github.com/fgnt/nara_wpe}}, which is based on the MIT licence. 
The results are as shown in Table \ref{tab:wpe} for TDNN, TDNN + LSTM and TDNN-F systems. Systems with ``Aug" as ``Y" use 
{\em{\aug} data augmentation} with noises from CHiME-5.
We see that applying dereverberation improves the results for all the systems.
Since the Nara WPE implementation \cite{drude2018nara-wpe} 
has a more open license, we include it in our improved Kaldi recipe.

\begin{table}[htb]
  \centering
  \caption{\label{tab:wpe}WER(\%) results comparing methods for dereverberation}
  \begin{tabular}{l|c|c|c|c}
    System & Aug & No WPE & NTT & Nara \\
    \hline
    \hline
    TDNN & N & 78.8 & 78.4 & 78.2 \\
    TDNN & Y & 74.8 & 74.4 & 73.9 \\
    TDNN + LSTM & Y & 74.6 & 74.2 & 74.0 \\
    TDNN-F & Y & 72.3 & \bf 71.7 & \bf 71.7
  \end{tabular}
\end{table}

\subsection{I-vector improvement using 2-stage decoding}
\label{ssec:2stage-results}

In this section, we investigate the effect of using the 2-stage decoding
described in Section \ref{ssec:ivectors} 
for the various neural network architectures
described in Section \ref{ssec:nnet_architectures}. 
We train the networks using 
``worn + 400k'' dataset, which is augmented using {\em{\aug} data augmentation method} with additive
noises from CHiME-5 data. We also apply Nara WPE-based dereverberation at test time. 
From the results, we see that 2-stage decoding 
gives around 2\% absolute improvement in WER
in all the cases. We also find that the best
network architecture is the TDNN-F, which 
we set as the new baseline in our updated Kaldi recipe.

\begin{table}[htb]
  \centering
  \caption{\label{tab:2-stage} WER(\%) results using 2-stage decoding}
  \begin{tabular}{l|c|c}
    System & 1-stage & 2-stage \\
    \hline\hline
    TDNN & 73.9 & 71.9 \\
    TDNN + LSTM & 74.0 & 72.3 \\ 
    CNN + TDNN + LSTM & 71.9 & 70.6 \\
    \bf TDNN-F & \bf 71.7 & \bf 69.4
  \end{tabular}
\end{table}



\section{Conclusions}

In this paper, we presented the acoustic modeling efforts at JHU
for the CHiME-5 ASR system. We explored 
data augmentation approaches for improving robustness
and found that the {\em{\aug} data augmentation} \cite{ko2017augmentation} method using a
publicly available RIR generator is very effective. 
We used it with additive noises extracted from 
CHiME-5 training data to improve WER by 4\% absolute. 
We got an additional 0.5\% improvement using
speech dereverberation implemented in the publicly available Nara WPE tool. 
We investigated a 2-stage decoding approach, where the first stage is 
used to select reliable single-speaker frames for extracting i-vectors. 
This approach gave us a consistent 2\% improvement in WER. 
We explored various neural network architectures, 
and got another 2\% improvement using the factorized form of TDNN (TDNN-F). 
Overall, we achieved a WER improvement of 11.7\% over the 
baseline of 81.1\% released by the CHiME-5 challenge organizers 
using an ASR system comprising of a single acoustic model, a single language model and a single front-end enhancement, and built using publicly available tools. 
We finally release this improved system
as a new CHiME-5 Kaldi baseline recipe.





\bibliographystyle{IEEEbib}
\bibliography{strings,refs}

\end{document}